\documentclass{ws-ijmpd}
\usepackage{tipa}
\usepackage{amsmath}
\usepackage{txfonts}
\usepackage{stmaryrd}
\usepackage{amssymb}

\begin{document}

\markboth{Renxin Xu}{Can cold quark matter be solid?}

%
\catchline{}{}{}{}{}
%

\title{Can cold quark matter be solid?}

\author{Renxin Xu}

\address{School of Physics and State Key Laboratory of Nuclear
         Physics and Technology,\\
         Peking University, Beijing 100871, P. R. China;
         {\tt r.x.xu@pku.edu.cn}}

\maketitle


\begin{abstract}
The state of cold quark matter really challenges both
astrophysicists and particle physicists, even many-body physicists.
It is conventionally suggested that BCS-like color superconductivity
occurs in cold quark matter; however, other scenarios with a ground
state rather than of Fermi gas could still be possible. It is
addressed that quarks are dressed and clustering in cold quark
matter at realistic baryon densities of compact stars, since a
weakly coupling treatment of the interaction between constituent
quarks would not be reliable. Cold quark matter is conjectured to be
in a solid state if thermal kinematic energy is much lower than the
interaction energy of quark clusters, and such a state could be
relevant to different manifestations of pulsar-like compact stars.
\end{abstract}

\keywords{Dense matter; Pulsars; Elementary particles.}

\vspace{0.5cm}

First of all, I would like to note that the word ``{\em solid} '' in
the title does not relate to solid evidence for quark matter, but
represents a new {\em solid} state of quark matter. To identify a
quark star should certainly be a milestone and could be possible in
the future, but now there isn't any solid and model-independent
evidence yet.

The state of cold quark matter is debated since 1970s, being
interested by either particle physicists or astrophysicists.
Because of asymptotic freedom, extremely dense and cold matter is
supposed to be of a Fermi gas of free quarks, and a condensate of
quark pairs near the Fermi surface (i.e., color superconductivity;
CSC) may occur according to perturbative quantum chromodynamics
(pQCD) or QCD-based effective models. Astrophysically, however,
realistic cold quark matter could only exist in pulsar-like compact
stars, and those QCD-based speculations should be tested by
different manifestations of such compact stars.
I will explain why a solid state of realistic cold quark matter
would be necessary from a view point of astrophysics in \S1 and \S2.
More issues related are addressed in later sections.

\vspace{0.2cm}
\noindent%
{\bf 1. What if pulsars are neutron stars?}

What's the nature of pulsars (PSRs)? The final answer to the
question surely depends on the understanding of non-pertubative QCD,
and relates to one of the 7 Millennium Prize Problems named by the
Clay Mathematical Institute. Nevertheless, the models for pulsars
can be classified into 4 kinds: hadronic stars (no quark matter),
hybrid stars (quark matter in the cores), crusted and bare quark
stars (quark matter dominates). The former two are usually called as
neutron stars (NSs), while the latter two quark stars (QSs).
It is worth noting that hard and model-independent evidence to
identify a QS may only be relevant to {\em bare} QSs because their
quark surfaces have sharp difference distinguishable from
others\cite{xu02}.

Although it is still a matter of debate whether pulsars are neutron
or quark stars, some individual PSR-like stars with mass of $\sim
1M_\odot$ and $\sim 10$ km in radius are certainly detected.
Historically, pulsars were supposed to be associated with
oscillations of white dwarfs or NSs\cite{psr68}, but soon recognized
as spinning compact NSs\cite{Gold68}, the Kepler frequency of which
in Newtonian gravity is
$$
\nu_{\rm Kepler}=\sqrt{{G\over 3\pi}{\bar \rho}}\simeq
841\sqrt{{\bar \rho}\over 10^{14}{\rm g/cm^3}}~{\rm Hz}.
$$
The average densities of NSs or QSs, $\bar \rho$, are order of
$M_\odot/(4\pi (10~{\rm km})^3/3)\simeq 5\times 10^{14}$ g/cm$^3$, a
few nuclear densities, which could be high enough to satisfy
observational frequency $\nu< \nu_{\rm Kepler}$.
More than 40 years later, can the NS model works all the way when
more and more new phenomena of PSR-like stars are discovered?

1.1, {\em Isolated PSR-like stars: why non-atomic thermal spectra?}
Many theoretical calculations, first developed by
Romani\cite{Romani87}, predicted the existence of atomic features in
the thermal X-ray emission of NS (or crusted QS) atmospheres, and
advanced facilities of {\it Chandra} and {\it XMM-Newton} were then
proposed to build for detecting those lines in order to constrain
stellar mass and radius by spectral red shift and pressure
broadening.
However, unfortunately, none of the expected spectral features has
been detected with certainty up to now, and this negative test may
hint a fundamental weakness of the NS models.
Though conventional NS models cannot be ruled out by only non-atomic
thermal spectra since modified NS atmospheric models with very
strong surface magnetic fields\cite{HL03}\cdash\cite{TZD04} might
reproduce a featureless spectrum too, a natural suggestion to
understand the general observation is that pulsars are actually bare
QSs\cite{xu02} because of no atom there on the surfaces.

More observations, however, did show absorption lines of PSR-like
stars, particularly from an interesting one, 1E 1207, at $\sim 0.7$
keV and $\sim 1.4$ keV. When discovered, these lines were suggested
to be associated with the atomic transitions of once-ionized helium
in an atmosphere with a strong magnetic field, but thought to
require artificial assumptions as cyclotron lines\cite{SPZT02}.
This view was soon criticized by Xu, Wang and Qiao\cite{xwq03}, who
addressed that all the 4 criticisms\cite{SPZT02} about the cyclotron
mechanism can be circumvented and emphasized that 1E 1207 could be a
bare QS with surface field of $\sim 10^{11}$ G.
Further observations of both spectra feature\cite{Big03} and precise
timing\cite{GH07} favor the electron-cyclotron model of 1E 1207.
The bare QS idea may survive finally if other absorption features
(e.g., of the spectra of soft Gamma-ray repeaters and anomalous
X-ray pulsars) are also be cyclotron-originated.

1.2, {\em Isolated PSR-like stars: real small X-ray radiation
radii?}
One of the key differences between NSs and (bare) QSs lies in the
fact that NSs are gravitationally bound while QSs not only by
gravity but also by additional strong interaction due to the strong
confinement between quarks. This fact results in an important
astrophysical consequence that bare QSs can be very low mass with
small radii (and thus spinning very fast, even at sub-millisecond
periods\cite{Xu07a,DXQH09}), while NSs cannot.
We see in Fig.~1 that the radii of gravitationally bound QSs are
smaller than that of QSs in flat space-time. The radius difference
between QSs without and with gravity represents the power of
gravitational interaction, which is certainly strong as stellar mass
($M$) increases. It is evident from Fig.~1 that gravity cannot be
negligible when QS's mass $M\gtrsim (10^{-3}\sim 10^{-2}) M_\odot$
in phenomenological models\cite{LX09} of quark matter.
\begin{figure}[t]
\centerline{\psfig{file=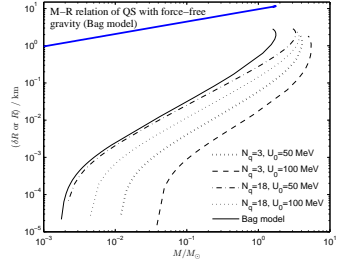,width=5cm}} \vspace*{8pt}
\caption{%
The radius difference ($\delta R$) between QSs without and with
gravity interaction, as a function of stellar mass. This differences
are correlated with the gravity in QSs. The mass($M$)-radius($R$)
relation of quark star in flat space-time is of $M\propto R^3$, that
is the straight thick line at the top of the figure. The curves of
$M-\delta R$ are for different models: solid line in a simple bag
model while others in Lennard-Jones quark matter models. The baryon
number density on the surface is chosen as two times of nuclear
density here.
({\em Data provided by Dr. Xiaoyu Lai})
\label{f1}}%
\end{figure}

Are there any observational hints of low-mass QSs?
Thermal radiation components from some PSR-like stars are detected,
the radii of which are usually much smaller than 10 km in blackbody
models where one fits spectral data by Planck
spectrum\cite{pavlov04}. Recently, Pavlov and Luna\cite{PL09} find
no pulsations with periods longer than $\sim 0.68$ s in the central
compact object (CCO) of Cas A, and constrain stellar radius and mass
to be \{$R=(4\sim 5.5)$ km, $M\lesssim 0.8M_\odot$\} in hydrogen NS
atmosphere models.
Two kinds of efforts are made toward an understanding of the fact in
conventional NS models. (1) The emissivity of NS's surface isn't of
blackbody or of hydrogen-like atmospheres. The CCO in Cas A is
suggested to covered by a carbon atmosphere\cite{HH09}. However, the
spectra from some sources (e.g., RX J1856) are still puzzling, being
well fitted by blackbody, especially with high-energy tails
surprisingly close to Wien's Formula: decreasing exponentially
($\propto e^{-\nu}$). (2) The small emission areas would represent
hot spots on NS's surfaces, i.e., to fit the X-ray spectra with at
least two blackbodies, but this has three points of weakness in NS
models. $a$, about $P$ and $\dot P$. No pulsation has detected in
some of thermal component-dominated sources (e.g., the Cas A
CCO\cite{PL09}), and the inferred magnetic field from $\dot P$ seems
not to be consistent with the atmosphere models at least for RX
J1856\cite{KK08}. $b$, fitting of thermal X-ray spectra (e.g., PSR
J1852+0040) with two blackbodies finds two small emitting radii
(significantly smaller than 10 km), which are not yet
understood\cite{HG10}. $c$, the blackbody temperature of the entire
surface of some PSR-like stars are much lower than those predicted
by the standard NS cooling models\cite{LiXH05}, even provided that
hot spots exist.
Nevertheless, besides that two above, a {\em natural} idea could be
that the detected small thermal regions ({\em if} being global) of
CCOs and others may reflect their small radii (and thus low masses
in QS scenario\cite{xu05mn}).

How can low-mass QSs be created?
Low-mass QSs are supposed to form during AIC (accretion-induced
collapse) of white dwarfs (WD)\cite{xu05mn}.
For a WD approaching the Chandrasekhar limit, with mass of $M_{\rm
wd}\sim 1.4M_\odot$ and radius of $R_{\rm wd}\sim 10^8$ cm, it's
gravitational energy is $E_{\rm g}\sim (3/5)GM_{\rm wd}^2/R_{\rm
wd}\simeq 3\times 10^{51}$ ergs.
If the energy release during detonation combustion of hadronic
matter to strange quark matter, starting from near the stellar
center, is responsible for the WD exploding, a necessary QS's mass
($M_{\rm qs,min}$) should satisfy $0.1 M_{\rm qs,min}c^2\simeq
E_{\rm g}$, where $10\%$ of rest mass is liberated, corresponding to
$\sim 100$ MeV per baryon. We then have $M_{\rm qs,min}\simeq
2\times 10^{-2}M_\odot$ (i.e., for a QS with radius of $\sim 2$ km).
Certainly such a QS should be bare after photon-driven explosion,
and should keep to be bare if it hasn't a history of Super-Eddington
accretion\cite{xu02,xu06}.
An accreted ion (e.g., a proton) should have enough kinematic energy
to penetrate the Coulomb barrier of a QS with mass $M_{\rm qs}$ and
radius $R_{\rm qs}$, as long as $GM_{\rm qs}m_p/R_{\rm qs}> V_q$.
The Coulomb barrier, $V_q$, is model-dependent, which varies from
$\sim 20$ MeV to even $\sim 0.2$ MeV. Approximating $M_{\rm
qs}=(4/3)\pi R_{\rm qs}^3\rho$ for low-mass QSs, we have a necessary
condition of
$$
M_{\rm qs}>\sqrt{3V_q^3\over 4\pi G^3 m_p^3\rho} \simeq 6\times
10^{-4} V_{q1}^{3/2} \rho_2^{-1/2} M_\odot,
$$
where the density $\rho=\rho_2\times (2\rho_0)$, with $\rho_0$ the
nuclear density, and $V_q=V_{q1}$ MeV.
In a word, strange QSs with mass as low as $\sim 10^{-2}M_\odot$
could form and would keep them bare if they are without strong
accretion history.

1.3, {\em Radio pulsars: how to reproduce drifting sub-pulses in
pulsar magnetospheres?}
Although PSR-like stars have many different manifestations, they are
populated by radio pulsars. Abundant radio pulses are not applied to
constrain the state of dense matter until Xu {\em et
al.}\cite{xq98,xqz99} addressed that radio pulsars could be bare QSs
and that the bare quark surface provides peculiar boundary
conditions for pulsar's magnetosphere electrodynamics. Additionally,
this certainly opens a new window to distinguish QSs from
conventional NSs via their magnetospheric activities.

Among the magnetospheric emission models for pulsar radio radiative
process, the user-friendly nature of Ruderman-Sutherland\cite{rs75}
model is a virtue not shared by others, and clear drifting
sub-pulses (especially bi-drifting) suggest the existence of
gap-sparking in the model.
However, that model can only work in strick conditions: strong
magnetic field and low temperature on surfaces of pulsars with ${\bf
\Omega \cdot B}<0$, while calculations showed, unfortunately, that
these conditions usually cannot be satisfied there.
This problem might be alleviated within a partially screened
model\cite{gm06} for NSs with ${\bf \Omega \cdot B}<0$, but could be
naturally solved for any ${\bf \Omega \cdot B}$ in the bare QSs
scenario.

1.4, {\em Birth of PSR-like stars: can supernova be successful?}
It is still an unsolved problem to simulate supernovae successfully
in the neutrino-driven explosion models of NSs.
Nevertheless, in the QS scenario, the {\em bare} quark surfaces
could be essential for successful explosions of both core and
accretion-induced collapses\cite{xu05mn}. The reason is that,
because of the strong binding of baryons, the photon luminosity of a
quark surface is not limited by the Eddington limit, and it is thus
possible that the prompt reverse shock could be revived by
photons\cite{orv05,cyx07}, rather than by neutrinos.

\vspace{0.2cm}
\noindent%
{\bf 2. What if pulsars are quark stars?}

Although it could be an attractive idea to solve the problems listed
in \S1 in the QS scenario, can we understand all of the different
manifestations in QS models? The answer could be ``yes''.
In principle, the discrepancies between observations and
expectations in previous QS models (e.g., the bag model) could be
explained if we conjecture a {\em new} solid state of cold quark
matter in compact stars.
I will summarize those issues in this section, demonstrating that
the solid state is very necessary.

2.1, {\em Thermal spectra: why Planck-like?}
There is certainly no atomic absorption in a bare QS's thermal
spectrum, but can the spectrum be well described by Planck's
radiation law?
In bag models where quarks are nonlocal, one limitation is that bare
QSs are generally supposed to be poor radiators in thermal X-ray
because of their high plasma frequency, $\sim 10$ MeV.
Nonetheless, if quarks are localized to form quark-clusters in cold
quark matter due to very strong interactions, a regular lattice of
the clusters (i.e., similar to a classical {\em solid} state)
emerges as a consequence of the residual interaction between
clusters\cite{xu03}.
In this latter case, the metal-like solid quark matter would induce
a metal-like radiative spectrum, with which the observed thermal
X-ray data of RX J1856 can be fitted\cite{zxz04}.
Exact emissivity of such solid quark matter cannot be calculated now
because of non-perturbative QCD, but that does not means that one
should not pursue this idea before finishing a QCD-based
calculation.
Alternatively, other radiative mechanism in the electrosphere (e.g.,
electron bremsstrahlung in the strong electric
field\cite{Zakharov10}) may also reproduce a Planck-like spectrum.

2.2, {\em Radio pulsars: normal and slow glitches}.
A big disadvantage that one believes pulsars are QSs lies in the
fact that the observation of pulsar glitches conflicts with the
hypothesis of conventional QSs in fluid states\cite{Alpar87,BHV90}
(e.g., in MIT bag models).
That problem could be solved in a solid QS model since a solid
stellar object would inevitably result in star-quakes when strain
energy develops to a critical value. Huge energy should be released
(and thus large spin-change occurs) after a quake of a solid quark
star because of the almost homogenous distribution of density.
Star-quakes could then be a simple and intuitional mechanism for
pulsars to have glitches frequently with large amplitudes.
In the regime of QSs, by extending the model for normal
glitches\cite{z04}, one can also model pulsar's slow
glitches\cite{px07} not to be well understood in NS models.

2.3, {\em Exploding events: AXPs/SGRs and GRBs.}
Solid QSs can have substantial free energy, both elastic and
gravitational, to be released after star-quakes, which would power
some extreme events detected in anomalous X-ray pulsars (AXPs) and
soft $\gamma$-ray repeaters (SGRs) and during $\gamma$-ray bursts
(GRBs).
Besides persistent pulsed X-ray emission with luminosity well in
excess of the spin-down power, AXPs/SGRs show occasional bursts
(associated possibly with glitches), even superflares with isotropic
energy $\sim 10^{44-46}$ erg and initial peak luminosity $\sim
10^{6-9}$ times of the Eddington one.
They are speculated to be {\em magnetars}, with the energy reservoir
of magnetic fields $\gtrsim 10^{14}$ G (to be still a matter of
debate about the origin\cite{Spruit08} since the dynamo action might
not be so effective and the strong magnetic field could decay
effectively), but could be solid quark stars with surface magnetic
fields similar to that of radio pulsars. Star-quakes are responsible
to both bursts/flares and glitches in the latter
scenario\cite{Xu07b}.
The most conspicuous asteroseismic manifest of solid phase of quark
stars is their capability of sustaining torsional shear oscillations
induced by SGR's starquake\cite{Sergey09}.
In addition, there are more and more authors who are trying to
connect the GRB central engines to SGRs' flares in order to
understand different GRB light-curves observed, especially the
internal-plateau X-ray emission\cite{xl09}. Besides the energy
released during deconfinement phase transition\cite{bpv04}, extra
ones are liberated after quakes.

2.4, {\em Free or torque-induced precession}.
Rigid body precesses naturally, but fluid one can hardly. The
observation of a few precession pulsars may suggest a totally solid
state of matter.
As is shown in Fig.~1, low-mass QSs with masses of $\lesssim
10^{-2}M_\odot$ and radii of a few kilometers are gravitationally
force-free, and their surfaces could then be irregular,
asteroid-like. Therefore, free or torque-induced precession may
easily be excited and expected with larger amplitude in low-mass
QSs.
The masses of AXPs/SGRs are approaching the mass-limit ($>M_\odot$)
in the AIQ model\cite{Xu07b}, they could then manifest no or weak
precession as observed, though they are more likely than CCOs/DTNs
(eg., RX J1856) to be surrounded by dust disks because of their
higher masses (thus stronger gravity).

\vspace{0.2cm}
\noindent%
{\bf 3. Could cold quark matter be solid?}

Due to QCD's asymptotic freedom, cold dense quark matter would
certainly be of Fermi gas or liquid if the baryon density is
extremely high, with a quark chemical potential $\sim 0.4$ GeV for
typical QSs (mass $\sim 1.4M_\odot$ and radius $\sim 10$ km). Most
of physicists are then basing their researches on this Fermi matter.
However, the problem is: Can the potentials be high enough so that
the interaction between quarks is negligible?
We may have a negative answer presented in this section, and
previous researches show also that pQCD would work reasonably well
only for quark chemical potentials above 1 GeV at least.

We note that the strong interaction between quarks in compact stars
may result in the formation of quark clusters, with a length scale
$l_q$ and an interaction energy $E_q$.
An estimate from Heisenberg's relation gives if quarks are
dressed\cite{xu09}, with mass $m_q\simeq 300$ MeV,
$$%
l_q \sim {1\over \alpha_s} {\hbar c\over m_qc^2}\simeq {1\over
\alpha_s}~{\rm fm},~~~ E_q \sim \alpha_s^2m_qc^2\simeq
300\alpha_s^2~{\rm MeV}.
$$%
This is dangerous for the Fermi state of matter since $E_q$ is
approaching and even greater than the potential $\sim 400$ MeV if
the running coupling constant $\alpha_s>1$.
With the estimation about $\alpha_s$ from recent work on
perturbative\cite{running} and non-perturbative\cite{dse1,dse2} QCD,
we can draw a numerical coupling as function of baryon number
density, shown in Fig.~2, assuming the energy scale likely of order
the chemical potential $\sim \hbar c (3\pi^2)^{1/3}\cdot n^{1/3}$
MeV ($n$: quark number density).
\begin{figure}[t]
  \centering
    \includegraphics[width=5cm]{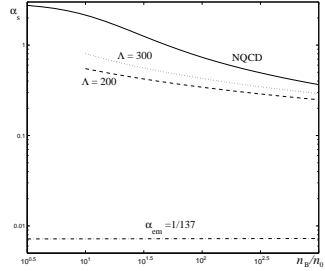}
    \caption{%
The running coupling, $\alpha_{\rm s}$, in cold quark matter as a
function of baryon density, $n_{\rm B}$. Both perturbative (for
cut-off parameter $\Lambda=200$, and 300) and non-perturbative QCD
(solid line, ``NQCD'') calculations are presented, respectively. For
comparison, the electromagnetic coupling of $\alpha_{\rm em}$ is
also drawn. The coupling would be $\alpha_{\rm s}\gtrsim 2$ in cold
quark matter at a realistic baryon density ($\sim$ a few nuclear
density, $n_0$).
\label{coupling}}
\end{figure}
At a few nuclear density in compact stars, the color coupling should
be very strong rather weak and we may have $\alpha_{\rm s}\gtrsim 2$
if non-perturbative QCD effects are included.
This surely means that a weakly coupling treatment could be
dangerous for realistic cold quark matter, and quarks would be
clustered and localized there.

In the QCD phase diagram, at extremely unrealistic high density,
cold quark matter would be of Fermi gas, and condensation in
momentum space may occur near the Fermi surface due to color
interaction (a BCS-like color super-conducting state). However, as
density decreases, various of interesting phases of quark matter may
appear. Firstly, as interaction strength increases, quarks may
condensate in position space to form different kinds of Bosons, and
Bosons may condensate (a BEC state) at low temperature. This is
called as BCS-BEC crossover. Secondly, a much stronger coupling
between quarks may favor Bosons to condensate in position space and
quark clusters form. This quark cluster matter could be in a liquid
(solid) state at high (low) temperature.

\vspace{0.2cm}
\noindent%
{\bf 4. Conclusions and Discussions}

The nature of PSR-like stars depends on the physics of cold matter
at supra-nuclear density, which is related to one of the challenging
problems nowadays: the non-perturbative QCD.
Besides efforts tried in QCD or QCD-based models from first
principles, terrestrial experiments and astronomical observations
can also provide valuable information of cold dense matter.
It is conjectured from an astrophysical point of view that cold
quark matter in compact star is in a solid state and PSR-like stars
are actually solid QSs.
We find that this conjecture would be correct if color interaction
between dresses quarks is still very strong, with a coupling
constant $\gtrsim 1$, in realistic cold quark matter.

That conjecture may have significant implications for the
fundamental strong interaction.
An essential point we proposed is that quarks should be dressed
(i.e., the chiral symmetry is broken) when density and temperature
are marginally high enough that hadronic degree of freedom freezes
while quark degree begins to free. This means that quarks have QCD
masses, rather than only Higgs masses, at an energy scale of $\sim
10^2$ MeV.
Hadronization in the early Universe would accordingly not be
possible due to the strong color interaction {\em if} the
electro-magnetic interaction isn't included (i.e., the charge of
quark could be negligible).
Nucleon is the lightest, but electrons with typical energy
$\mu_e\sim 10^2$ MeV have to participate in nuclear matter. Such
nuclear matter would be unstable, to evaporate into a nucleon gas
with electron's kinetic energy $\ll \mu_e$. Big Bang nucleosynthesis
occurs then when the temperature of nucleon gas cools.
Nevertheless, strange quark nuggets should form during cosmic QCD
phase transition because of a very low charge-mass ratio and a high
binding energy per baryon.
Additionally, multi-quark clusters may temporarily form in nuclei in
order to understand the puzzling EMC effect\cite{CH83,EMC03}.

We try to raise a possibility of quark matter in a solid state, but
never to present a general review. We feel sorry for neglecting many
interesting references related.

\vspace{2mm}%
{\em Acknowledgments}:
I would like to thank Prof. Sergey Bastrukov and Prof. Marten van
Kerkwijk for valuable discussions, and to acknowledge various
contributions by members at the pulsar group of PKU.
This work is supported by NSFC (10973002, 10935001) and the National
Basic Research Program of China (2009CB824800).

\end{document}